\documentclass[twocolumn,prb,superscriptaddress,preprintnumbers,amssymb]{revtex4}
\usepackage{graphicx}
\usepackage{dcolumn}% Align table columns on decimal point
\usepackage{bm}% bold math

\begin{document}

\title
{ Electronic properties of the partially hydrogenated armchair carbon nanotubes}

\author{\v{Z}eljko \v{S}ljivan\v{c}anin}
\affiliation{Vin\v{c}a Institute of Nuclear Sciences (020),
University of Belgrade, RS-11001 Belgrade, Serbia}
\email{zeljko@vinca.rs}
\affiliation{Interdisciplinary Nanoscience Center (iNANO),
University of Aarhus, DK-8000 {\AA}rhus C, Denmark}

\date{\today}

\begin{abstract}
By means of pseudopotential calculations based on density functional theory (DFT) 
we studied the effect of hydrogenation on electronic properties of armchair single-wall 
carbon nanotubes. The calculations demonstrate strong preference for formation of monoatomic 
H chains along the (5,5) nanotube axis with the H binding in an infinite H chain reaching the value 
of 2.58 eV per atom. Upon formation of chains of H adatoms, initially metallic (5,5) nanotubes 
change electronic structure to the semiconducting. The opening of the band gap of $\sim$0.6 eV is 
accompanied with antiferromagnetic coupling of ferromagnetically ordered magnetic moments on C 
atoms in vicinity of the H chain. These electronic properties are strikingly similar to those 
previously observed in narrow graphene nanoribbons with zigzag edges.
\end{abstract}


\maketitle
\section{Introduction}\label{intro}
Graphene and graphene-based materials have attracted a great deal of interest due 
to their high stability and peculiar electronic properties, \cite{Novoselov2005,Geim2007,
Castro2009,Saito1998} which make them suitable candidates for sensing applications,
\cite{Schedin2007,Kong2000} nanoelectronic components, 
\cite{Avouris2007,Li2008,McEuen2002} 
or media for hydrogen storage. \cite{Dillon1997,Zuttel2001,Baughman2002} The most studied 
aspect of graphene is its electronic properties. \cite{Castro2009} At variance to 
conventional semiconductors, the electrons in graphene can move ballistically even at room 
temperature, which means the graphene can be used for design of electronic devices operating 
at the frequencies beyond the limits of the electronics based on silicon. 
However, being a zero-band-gap semiconductor graphane can not be used directly 
as a field-effect transistor for logic applications. 
Different strategies adopted in attempts to open a band gap in graphene include 
the substrate-induced band-gap, \cite{Zhou2007} the use of bilayer graphene, 
\cite{McCann2006,Oostonga2008} graphene cutting into nanoribbons (GNR), 
\cite{Son2006a,Son2006,Barone2006,Han2007} or its hydrogenation. 
\cite{Duplock2004,Sofo2007,Elias2009,Liv2010,Haberer2010} Only  graphene hydrogenation and 
fabrication of GNR are methods able to open the band gap of 0.4 eV or more, as required for the 
applications in electronic devices operating at room temperature. However, hydrogen 
functionalization has been realized only on a conducting substrate, namely Ir(111). \cite{Liv2010}
This approach cannot be easily transferred to graphene on insulating 
substrates, which is needed for applications in electronics.\\ 
The GNRs have emerged as structures with great potential to become building blocks for carbon-based 
nanoelectronics after prediction to have band gaps useful for room temperature transistors,
excellent switching characteristics, and ballistic transport properties. 
\cite{Son2006,Avouris2007,Li2008}  At variance to carbon nanotubes (CNTs) which exhibit 
extreme chirality dependence of metallic or semiconducting nature, the GNRs with the width below 
10 nm are all semiconductors. \cite{Son2006a,Barone2006} Yet, producing narrow GNRs with high 
quality edges in order to preserve high-electron mobility of graphene remains a big challenge for 
state-of-the-art  lithographic, chemical, or sonochemical methods. Recently, Jiao {\it et al.} 
\cite{Jiao2009} developed a new route toward fabrication of 
the GNRs with smooth edges, starting from CNTs with narrow diameter and chiral distribution.
Since the methods for synthesis and size control of the CNTs are well established, Jiao and 
co-workers based the fabrication of  high-quality GNRs on unzipping of carbon nanotubes. 
The difficult task of cleaving CNTs along the axis was achieved by an Ar plasma etching method. 
The GNRs with smooth edges and width in range from 10  to 20 nm were produced. Further 
improvements of the method were realized through the mechanical sonification of mildly oxidized 
multiwall carbon nanotubes. \cite{Jiao2010} \\
%
% Hydrogenation
%
The decoration of graphene with H monoatomic lines was recently proposed in the 
computational studies of Chernozatonskii {\it et al.} \cite{Chernozatonskii2007,Chernozatonskii2010}  
as a new method to modify the electronic structure of graphene. The authors demonstrated that periodic 
H lines divide graphene into electronically independent strips which show electronic properties close 
to those of GNRs with armchair edges. Yet, the method proposed for the fabrication of these 
structures on graphene is questionable. 
Miller {\it et al.} \cite{Miller2008} combined experimental observations with computational methods to 
investigate the hydrogenation of CNTs using high boiling polymers as hydrogenation reagents.
The measured values of the C-H stretching vibrations are in good agreement with the those obtained for 
CNTs hydrogenated with H plasma. \cite{Khare2002}
According to the first-principles calculations reported in Ref. \cite{Miller2008} the chemisorbed 
hydrogens should preferentially form axially aligned chains.\\
In the present work we focus on the electronic structure of hydrogenated CNTs and provide clear
evidence that the electronic properties of metallic armchair CNTs can be tuned to those of semiconducting 
GNRs through adsorption of H atoms, without need for unzipping of CNTs. 
Our {\it ab-initio} calculations reveal that partial hydrogenation of CNTs leads to the formation 
of H chains along the tube axis, the structures which are thermodynamically favorable and 
kinetically stable at room temperature. We demonstrate that the CNTs decorated with H chains are 
magnetic semiconductors with electronic properties remarkably similar to those of GNRs with 
zigzag edges.  These results would contribute to development of experimental methods in band-gap 
engineering of carbon nanostructures based on their hydrogenation rather than high precision 
cutting of graphene or carbon nanotubes.\\
\section{Computational methods}\label{comp}
The DFT calculations were performed with the plane-wave-based DACAPO program
package, \cite{Hammer1999,Bahn2002} applying ultra soft pseudopotentials
\cite{Vanderbilt1990,Laasonen1993} to describe electron-ion interactions,
and the Perdew-Wang functional (PW91) \cite{pw91} for the electronic exchange 
correlation effects. The electron wave functions and augmented electron density were
expanded in plane waves with cutoff energies of 25 Ry and 140 Ry,
respectively. In our investigation of the H interaction with metallic single-wall carbon 
nanotubes (CNTs) we have chosen the (5,5) nanotubes, since this class of extensively studied 
nanotubes offers the possibility to perform required calculations at modest computational 
cost. The calculations were performed using a supercell with dimensions of 19 {\AA}  in
directions perpendicular to the  nanotube axis and 9.84 {\AA} along the axis. 
The Monkhorst-Pack scheme  \cite{Monkhorst1976} with sixteen points was applied  for sampling of the Brillouin zone. 
The activation energies related to the H diffusion on the CNT(5,5) were calculated 
using the nudged elastic band method. \cite{neb} In order to demonstrate that the main results of the study 
are applicable to other metallic nanotubes we performed a limited number of calculations for the 
(8,8) and (10,10) CNTs. All H binding energies are given relative to the energy of the free H atom.\\
Since the electronic and magnetic properties of nanostructures are often 
sensitive to the choice of pseudopotentials and their implementation we 
performed a set of additional calculations with the Quantum Espresso (QE) package. 
\cite{QE} The tests demonstrated that the results obtained with the DACAPO and QE 
codes are in very good agreement. The H binding energies at CNT(5,5) differ 
less than 0.03 eV per atom; the calculated band-gaps for H wire on CNT(5,5) 
[Fig.\ref{clusters}d] are 0.58 eV (DACAPO) and 0.53 eV (QE). The local magnetic
moments as well as local electronic properties are nearly identical. Hence,
in the rest of the paper we present and discuss only results produced with
the DACAPO code.
\section{Results and Discussion}
\subsection{Structural properties of partially hydrogenated armchair CNTs}
\begin{figure}[th]
\includegraphics[width=1.0\linewidth]{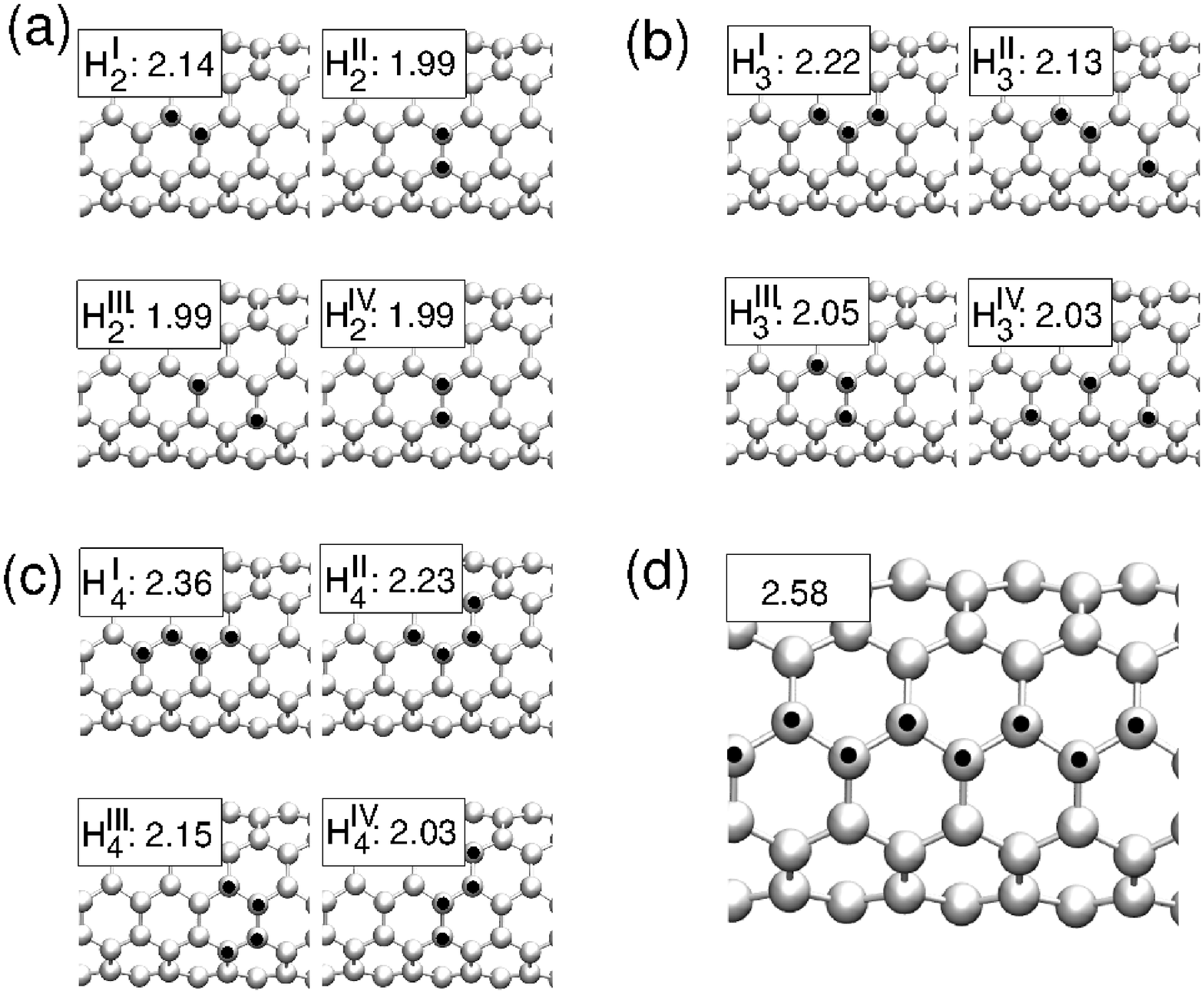}
\caption{\label{clusters} The configurations of (a) dimers, (b) trimers, (c) tetramers, and 
(d) an infinite chain of H adatoms on CNT(5,5). The corresponding binding energies per H atom are in 
eV. H and C atoms are represented with small black and gray spheres, respectively. 
}
\end{figure}

%
% Monomer
%
The initial step in our investigation of the CNT(5,5) hydrogenation was
the search for the thermodynamically most favorable structures formed by deposited H atoms.
Starting with H monomers we calculated a binding energy of 1.67 eV, which  is considerably 
higher than the value of 0.8 eV obtained for graphene, applying the same computational
approach. \cite{Zeljko2009} 
%
% Dimers
%
The H dimers are the smallest H clusters  on graphene stable at room temperature.
\cite{Liv2006} Several recent studies identified the ortho- (O)
and para-dimers (P) as the most favorable dimer structures with similar H binding energies.
\cite{Liv2006,Casolo2008,Zeljko2009}
The dimer adsorption picture on CNT(5,5) is more complicated than on graphene due to axial 
symmetry of the nanotube which leads to different types of the O and P dimers.  
The most stable dimer configuration  on CNT(5,5) identified from our calculations are  
shown in Fig. \ref{clusters}(a). 
Thermodynamically  preferential are tilted O dimers (configuration H$_2^I$), with binding energy 
of 4.28 eV, 0.94 eV more than binding of two isolated H atoms. This high binding energy 
confirms the strong tendency for clustering of H adatoms on carbon nanotubes, in agreement with
expectation based on results produced for H adsorption on graphene. \cite{Zeljko2009} Other 
favorable dimer structures depicted in Fig. \ref{clusters}(a) are 0.3 eV higher in energy than tilted 
O dimers. \\
According to our previous study \cite{Zeljko2011} the most stable configurations of H clusters 
with $n$ adatoms (2 $\le$ $n$ $\le$ 6) on graphene are exclusively composed of O and P dimers. 
The test calculations performed for H dimers and trimers on CNT(5,5) show that the same effect 
is found on armchair carbon nanotubes. This significantly reduces the number of H configurations 
which should be included in the search for the most stable H structures. Therefore, in the 
following  we consider only H configurations in which all neighboring H atoms were configured
as in O or P dimers.\\  
The calculations performed for H trimers provide evidence for tendency of H adatoms to cluster into 
chains oriented along the tube axis [see Fig. \ref{clusters}(b)]. The axially aligned H$_3^I$ 
structure is 0.24 eV more favorable than the H$_3^{II}$ trimer. The other two trimers in Fig. 
\ref{clusters}(b) are 0.51 eV (H$_3^{III}$) and 0.57 eV (H$_3^{IV}$) higher in energy  than the H$_3^I$ 
cluster. \\
%
% Tetramers
%
Four tetramers considered in our study are presented in Fig. \ref{clusters}(c).
The chain of H adatoms along the tube axis ( H$_4^{I}$) is 0.5 eV more stable than the H$_4^{II}$ 
structures. The third structure (H$_4^{III}$) is less stable by an additional
0.34 eV. The H$_4^{IV}$ cluster oriented perpendicular to the tube axis is as much as 1.32 eV
less stable than the  H$_4^{I}$ configuration, which further confirms the strong preference of H 
adatoms to align along the nanotube axis. \\
\begin{figure}[th]
\includegraphics[width=0.8\linewidth]{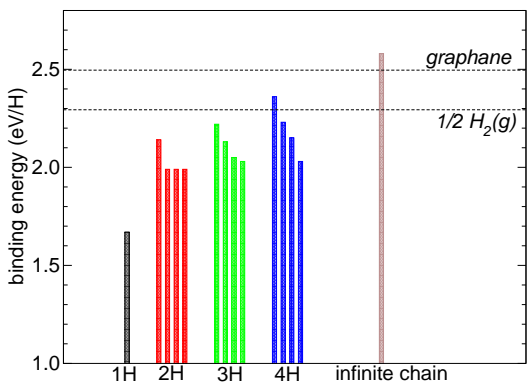}
\caption{\label{binding} Evolution of the H binding energy with the size of the 
configurations in Fig. \ref{clusters}. The calculated values are compared to those 
obtained in H$_2$(g) and graphane.
}
\end{figure}
An infinite chain of H atoms along the tube axis, depicted in Fig. \ref{clusters}(d), is the 
particularly stable adsorption configuration. In the rest of the text we refer to this structure as 
the H-CNT(5,5). The calculated H binding energy is 2.58 eV per atom, considerably more than the 
2.29 eV/H calculated for the gas-phase H$_2$ molecule or the 2.49 eV/H obtained for graphane. 
The chain of H adatoms was previously reported as a favorable configuration of H adatoms in a study 
of debundling and dispersing of CNTs upon their hydrogenation. \cite{Miller2008} 
The H binding energies of all structures presented in Fig. \ref{clusters} are compared in 
Fig. \ref{binding}. In addition to the observed preference for axially aligned H chains, 
the plot clearly demonstrates an enhancement in thermodynamic stability of the chains with 
increase in their size.\\ 
We conclude our description of the structural properties of H chains on CNT(5,5) by examining
their kinetic stability against H diffusion to the nearest C atoms, in the direction perpendicular to 
the nanotube axis. The calculated activation energy of 1.7 eV confirms that chains of H atoms 
aligned along the tube axis are kinetically highly stable at room temperature. \\
\subsection{Electronic properties of partially hydrogenated armchair CNTs}
\begin{figure}[ht]
\includegraphics[width=0.9\linewidth]{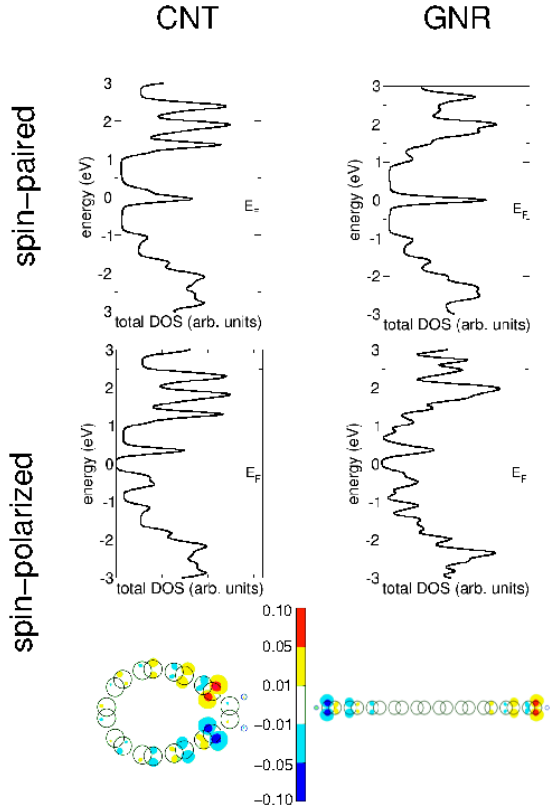}
\caption{\label{dos} Total density of states (DOS) of CNT(5,5) decorated with H chain 
and 10-ZGNR, produced from spin-paired (top panels) and spin-polarized calculations (middle panels), 
as well as the corresponding spin densities (lowest panel) averaged along the nanotube or 
nanoribbon axis. 
}
\end{figure}

Once the favorable structure of H adsorbates on the carbon nanotube is identified we focus 
on the electronic properties of the CNT(5,5) decorated with the chain of H adatoms.
It turns out that the adsorbates profoundly change the electronic structure of the CNT(5,5), 
transforming it from a metal to a magnetic semiconductor. The  $\pi$-bonding network of electronic 
bands in the vicinity of the Fermi level which originates from the 2$p_z$ states of individual C 
atoms can be disrupted through hydrogen adsorption, since the  contribution from the C 2$p_z$ 
orbitals of hydrogenated C atoms is removed from the  $\pi$ bonds near the Fermi level upon their  
hybridization with H 1$s$ orbitals. Hence, instead of cutting graphene into ribbons, the edges at 
the  $\pi$-bonding network of  the C-2$p_z$ 
states can be produced with chains of adsorbed H atoms. The zigzag chain of H adatoms breaks the 
cyclic boundary conditions of pristine CNT(5,5), creating open boundary conditions similar to those 
in zigzag carbon nanoribbons (ZGNRs). Thus, the electronic structure of CNT(5,5) with the H line is 
expected to show considerable similarity to the electronic structure of ZGNRs. We now illustrate the 
striking similarity of the valence and conduction bands in GNRs and hydrogenated CNT(5,5) by 
comparison of their main features:\\
\begin{figure}[ht]
\includegraphics[width=0.7\linewidth]{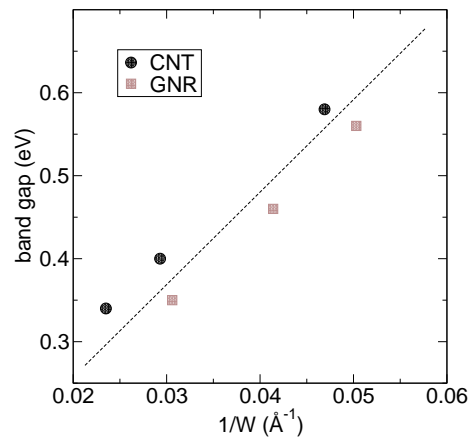}
\caption{\label{gaps}  Evolution of the band-gap size of partially hydrogenated CNTs and ZGNRs with
their width (W). We assume that the widths of CNTs are identical to their circumferences.
}
\end{figure}
{\it (i)} The calculated band gap in H-CNT(5,5) is 0.58 eV. Our calculations for bigger 
nanotubes indicate decrease of the band-gap size with the nanotube diameter. For H-CNT(8,8) and 
H-CNT(10,10) we calculated values of 0.40 and 0.34 eV, respectively. The trends in the band gaps  
of the H-CNTs are fully in line with those observed for ZGNRs, 
\cite{Son2006a,Barone2006} as demonstrated in Fig. \ref{gaps}, where we presented results for 
partially hydrogenated CNT(n,n) (n =5,8,10), as well as for m-ZGNRs (m=8,10,16), chosen as the 
nanostructures with widths comparable to those of the CNTs considered here. \cite{width} \\ 
{\it (ii)} The band-gap opening in H-CNTs and ZGNRs is driven by the same 
physical mechanism, i.e.  the exchange interaction. To demonstrate this finding we performed 
non-spin-polarized calculations for two considered types of nanostructures. The corresponding total 
densities of states (DOS) are depicted in Fig. \ref{dos}. 
For both systems, the H-CNT(5,5) and 10-ZGNR,
very high values of DOS at $E_F$ are obtained in spin-paired calculations. The corresponding 
electronic states  are removed from the Fermi level when the on-site exchange interaction is
switched on, which results in induced magnetic moments on C atoms, as well as in the band-gap opening.
This mechanism of the band-gap opening in graphene-based nanostructures was already reported by 
Son {\it et al.} in their study of ZGNRs. \cite{Son2006a}
The magnetic moments at C atoms belonging to the same zigzag chain along the tube or nanoribbon are 
ferromagnetically coupled, with different signs of magnetic moments for C atoms located on 
different sides of the H chain or near different zigzag edges. The spin density rapidly decays 
with the distance from the H chain or zigzag edge. Its main features are very similar for H-CNT(5,5) 
and 10-ZGNR (see lowest panel in Fig. \ref{dos}). \\
\begin{figure}[ht]
\includegraphics[width=0.9\linewidth]{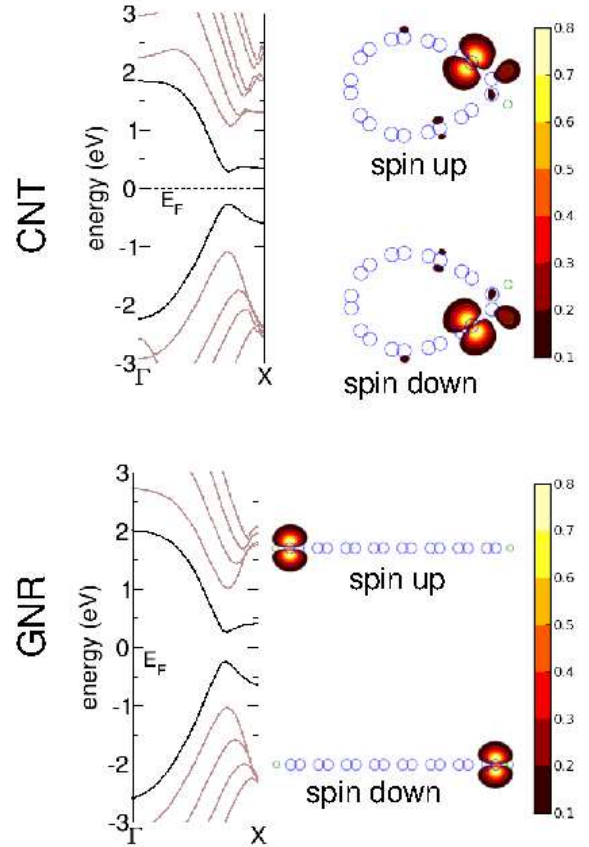}
\caption{\label{bands_wf} Band structure diagrams (left) and the absolute value of 
the Kohn-Sham states (right) of CNT(5,5) decorated with H chain and of 10-ZGNR.  The Kohn-Sham states 
of valence bands at X point are plotted in the planes perpendicular to the tube axis or nanoribbon edges. 
The H and C atoms are shown as small and big circles, respectively.
}
\end{figure}

\noindent
{\it (iii)} According to the tight-binding theory of ZGNRs \cite{Nakada1996,Brey2006} very high 
values of the DOS at Fermi level are due to  flat bands of the edge states. They are split due to 
exchange interaction and pushed below (valence band) and above (conduction band) $E_F$.
These states decay with the distance from the edge with the decay profile depending on the momentum: 
The highest localization at the edges is observed at the X-point. \cite{Klein1993,Fujita1996} 
The same type of edge states is also found in our study of the H-CNT(5,5). 
The edge states of H-CNT(5,5) and 10-ZGNR are compared in Fig. \ref{bands_wf}, where we 
presented isocontour plots of the absolute values of the Kohn-Sham states of the valence band at 
the X-point, for both spin orientations. The states corresponding to two different spin channels are 
located at different sides of the H chain [H-CNT(5,5)] or near different zigzag edges [GNRs].
The spin polarization of the edge states induces considerable magnetic moments on C atoms in 
the vicinity of the H lines (CNTs) or zigzag edges (GNR). The characters of these states in 
H-CNT(5,5) and 10-ZGNR (see Fig. \ref{bands_wf}) are remarkably similar. \\
\section{Conclusions}
Our thorough inspection of the electronic properties of armchair CNTs decorated with the 
chain of adsorbed H atoms demonstrates marked similarity between their electronic properties and 
those observed in ZGNRs. This similarity originates from the breaking of cyclic boundary 
conditions of pristine nanotubes upon H deposition, which leads to the creation of the edge 
states near the H chain, equivalent to the states observed at the edges of ZGNRs. Given the strong 
preference of H adatoms to assemble into chains oriented along the nanotube axis, together with 
the high kinetic stability of the formed structures, the partial hydrogenation of carbon nanotubes 
is a promising route for designing new nanostructures with electronic properties resembling those 
of narrow graphene nanoribbons. Our study indicates that the hydrogenation opens prospects for 
achieving electronic properties of narrow GNRs without utilization of the challenging techniques 
of cutting graphene sheets or unzipping carbon nanotubes with (sub)nanometric precision. \\
The comparison of the electronic structure of partially hydrogenated CNTs and 
GNRs should be extended to the study of the hydrogenation of zigzag CNTs which are 
expected to obey electronic properties similar to those of armchair GNRs. \\
\section{Acknowledgments}
We acknowledge valuable discussion with Filip\ R.\ Vukajlovi\'c.
This work has been supported by the Serbian Ministry of Education and Science 
under Grants No.\ 141039A and No.\ 171033. The calculations were performed at the Danish Center for 
Scientific Computing.


\begin{thebibliography}{10}
\bibitem{width}
We assume that the widths of CNTs are equavalent to their circumferences.
\end{thebibliography}
\end{document}